\documentclass[%
  floatfix,aip,
  amsmath,amssymb,
  reprint,
  ]{revtex4-1}

\usepackage[modulo,switch]{lineno}
\modulolinenumbers[1]
\usepackage{hyperref}
\usepackage[utf8]{inputenc}
\usepackage{graphicx}
\usepackage{amsmath}
\usepackage{amssymb}
\usepackage{bm}        % for math
\usepackage{txfonts}
\usepackage{multirow}
\usepackage{mathptmx}  % Times font

\usepackage{lipsum}                     % Dummytext
\usepackage{xargs}                      % Use more than one optional parameter in a new commands
\usepackage[pdftex,dvipsnames]{xcolor}  % Coloured text etc.
\usepackage[colorinlistoftodos,prependcaption,textsize=medium]{todonotes}
\newcommandx{\unsure}[2][1=]{\todo[inline,linecolor=red,backgroundcolor=red!25,bordercolor=red,#1]{#2}}
\newcommandx{\change}[2][1=]{\todo[inline,linecolor=blue,backgroundcolor=blue!25,bordercolor=blue,#1]{#2}}
\newcommandx{\info}[2][1=]{\todo[inline,linecolor=OliveGreen,backgroundcolor=OliveGreen!25,bordercolor=OliveGreen,#1]{#2}}
\newcommandx{\improvement}[2][1=]{\todo[inline,linecolor=Plum,backgroundcolor=Plum!25,bordercolor=Plum,#1]{#2}}
\newcommandx{\thiswillnotshow}[2][1=]{\todo[disable,#1]{#2}}

\begin{document}

\title{The Panopticon device: an integrated Paul-trap-hemispherical mirror system for quantum optics}

\author{G.~Araneda}
\altaffiliation{Corresponding author}
\email{gabriel.aranedamachuca@physics.ox.ac.uk}
\altaffiliation{Current address: Department of Physics, University of Oxford}
\affiliation{Institut f\"ur Experimentalphysik, Universit\"at Innsbruck, Technikerstrasse~25, 6020~Innsbruck, Austria}
\author{G.~Cerchiari}
\affiliation{Institut f\"ur Experimentalphysik, Universit\"at Innsbruck, Technikerstrasse~25, 6020~Innsbruck, Austria}
\author{D. B.~Higginbottom}
\affiliation{Department of Physics, Simon Fraser University, Burnaby, British Columbia, V5A 1S6, Canada}
\author{P. C.~Holz}
\affiliation{Institut f\"ur Experimentalphysik, Universit\"at Innsbruck, Technikerstrasse~25, 6020~Innsbruck, Austria}
\author{K.~Lakhmanskiy}
\affiliation{Institut f\"ur Experimentalphysik, Universit\"at Innsbruck, Technikerstrasse~25, 6020~Innsbruck, Austria}
\author{P. Ob\v{s}il}
\affiliation{Department of Optics, Palack\'{y} University, 17. listopadu 12, 771 46 Olomouc, Czech Republic}
\author{Y.~Colombe}
\affiliation{Institut f\"ur Experimentalphysik, Universit\"at Innsbruck, Technikerstrasse~25, 6020~Innsbruck, Austria}
\author{R.~Blatt}
\affiliation{Institut f\"ur Experimentalphysik, Universit\"at Innsbruck, Technikerstrasse~25, 6020~Innsbruck, Austria}
\affiliation{Institut f\"ur Quantenoptik und Quanteninformation, \"Osterreichische Akademie der Wissenschaften, Technikerstrasse 21a, 6020 Innsbruck, Austria}

\date{\today}

%%%%%%%%%%%%%%%%%%%%%%%%%%%%%%%%%%%%%%%%%%%%%%%%%
%%%%%%%%%%%%%%%%%%%%%%%%%%%%%%%%%%%%%%%%%%%%%%%%%

%Comments colorcoding \\
%\textcolor{red}{Giovanni} - \textcolor{blue}{Gabriel} - \textcolor{green}{Lukas P.}\\

\begin{abstract}
We present the design and construction of a new experimental apparatus for the trapping of single Ba$^+$ ions in the center of curvature of an optical-quality hemispherical mirror. We describe the layout, fabrication and integration of the
full setup, consisting of a high-optical access monolithic `3D-printed' Paul trap, the hemispherical mirror, a diffraction-limited in-vacuum lens ($\text{NA} = 0.7$) for collection of atomic fluorescence and a state-of-the art ultra-high vacuum vessel. This new apparatus enables the study of quantum electrodynamics effects such as strong inhibition and enhancement of spontaneous emission, and achieves a collection efficiency of the emitted light in a single optical mode of 31\,\%. 

\end{abstract}
\maketitle
\section{Introduction and motivation}
Atomic ions confined and laser-cooled in Paul traps provide an experimental platform to study quantum objects with a unique level of control granted by the localization of the atomic wavefunction over a few nanometers, the long trapping time and the high fidelity manipulation of internal and external degrees of freedom~\cite{leibfried2003quantum}. These features have been used extensively to build quantum information processors (see for example Ref.~\onlinecite{schindler2013quantum}), and for the study of fundamental phenomena in quantum optics and atomic physics (see for example Ref.~\onlinecite{appasamy1995quantum}). 

One of the most fundamental and intriguing phenomena in quantum optics is the spontaneous emission and the possibility to modify it~\cite{dirac1927quantum,welton1948some,kleppner1981inhibited,milonni1984spontaneous}. The spontaneous emission of atoms can be enhanced or inhibited in several ways. One way to modify the spontaneous emission rate is through dipole-dipole interactions between different atoms, namely sub- and superradiance-type effects~\cite{dicke1954coherence,gross1982superradiance,andreev1980collective,Benedict2006}. This has been extensively studied in neutral atom systems and used for diverse applications (see for example Refs.~\onlinecite{skribanowitz1973observation, van2003atomic,kuzmich2003generation,reimann2015cavity,chou2004single,black2005demand,chan2003observation,wolke2012cavity,bohnet2012steady}). In general it is difficult to observe large modifications of the spontaneous emission with trapped ions using this approach, since the Coulomb interaction restricts the minimum achievable inter-ion distance. Nevertheless, some degree of modification has been shown for the case of two ions. In Ref.~\onlinecite{devoe1996observation} enhancement of $\sim 1.5\,\%$ and inhibition of $\sim 1.2\,\%$ were observed by locating two ions 1470 nm away from each other. 

Another way to enhance or reduce the rate of spontaneous emission rate of an atom is to modify the electromagnetic vacuum mode structure interacting with the atom~\cite{purcell1946resonance,KIFFNER201085}. The electromagnetic mode structure can be altered by placing the atom close to dielectric interfaces~\cite{drexhage1970influence}, between two mirrors~\cite{haroche1989cavity} or inside photonic structures producing a bandgap~\cite{kweon1995quantum}. In particular, large enhancement of the emission from a single atom has been achieved using high-finesse cavities~\cite{brune1994lamb,heinzen1987vacuum,hood2000atom,kreuter2004spontaneous}. Recently, a five-fold enhancement in the spontaneous emission rate has been demonstrated by placing a single atom in a fiber cavity~\cite{gallego2018strong}. Furthermore, in the realm of solid state emitters, enhancement by a factor of more than 100 has been observed using microcavities~\cite{vahala2003optical}. In ion trap systems, strong coupling between a single trapped ion and a fiber cavity has been recently observed \cite{takahashi2020strong}. While a resonant cavity can increase the spontaneous emission rate into a particular mode, inhibiting the spontaneous emission requires to suppress the coupling to all modes. Only few experiments have demonstrated large inhibitions, see for example Ref.~\onlinecite{hulet1985inhibited} where the rate of spontaneous emission of a single Rydberg atom was reduced by a factor $\sim 20$, or Ref.~\onlinecite{bayer2001inhibition} where the emission rate of a solid state emitter was reduced by a factor $\sim 10$.

\begin{figure}
\centerline{\includegraphics[width=0.9\columnwidth]{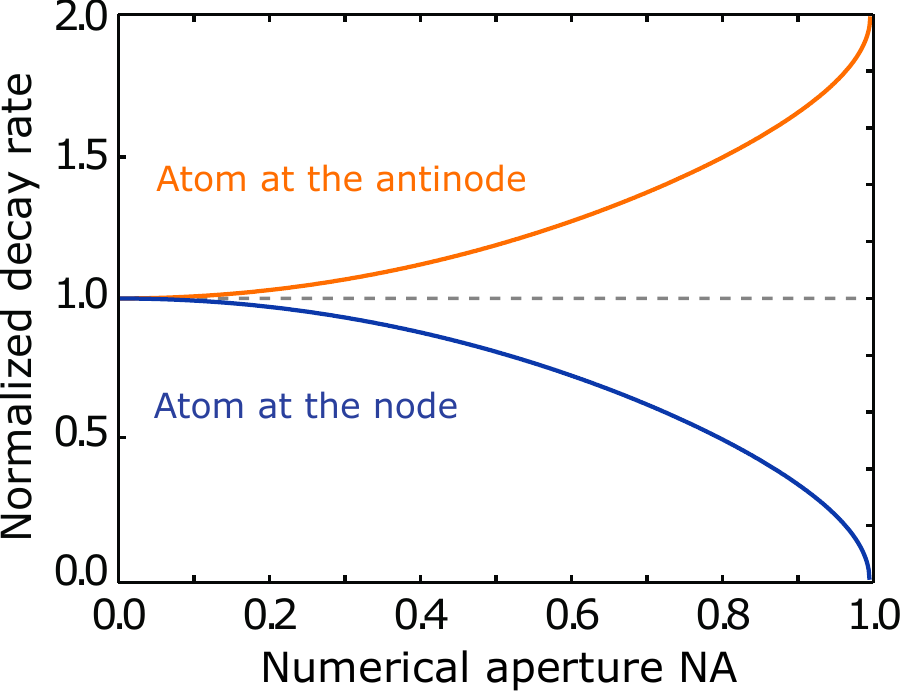}}
\caption[Modification of the spontaneous emission in a dipole transition of an atom in the center of a hemispherical mirror]{\textbf{Modification of the spontaneous emission rate of an atom in the center of a hemispherical mirror}. The spontaneous emission of an atom at the center of curvature of a spherical mirror is inhibited or enhanced depending on the radius and numerical aperture of the mirror. If the radius of the mirror is $R=n\lambda/2$, the center of curvature is a node of the vacuum mode density and inhibition is expected (blue curve). If the radius of the mirror is $R=n\lambda/2+\lambda/4$, the center of curvature is an anti-node of the vacuum mode density and enhancement is expected (orange curve). Full inhibition (resp. a doubling of the decay rate) is obtained when half of the space is covered by the mirror ($\text{NA} = 1$). In this plot a mirror with perfect reflectivity is assumed.
}
\label{fig:modification_emission}
\end{figure}

It is a common assumption in the field of cavity electrodynamics that full inhibition of the spontaneous emission can only be achieved by placing the atom between mirrors that cover the full solid angle around it, restricting all the vacuum modes resonant with the atom, see for example Ref.~\onlinecite{kleppner1981inhibited}. However, given the point symmetry of the atomic emission, it is possible to fully control the emission rate by covering only \emph{half} of the solid angle. This can be realized by placing the atom in the center of a concave hemispherical mirror, as theoretically predicted in Ref.~\onlinecite{hetet2010qed, hetet2020err}. Taking this approach, both inhibition and enhancement of the spontaneous emission can be achieved depending on the distance $R$ between the atom and the surface of the mirror, provided that the mirror is close enough to the atom to allow temporal interference of the field emitted in opposite directions. This condition can be written as $2R/c \ll 1/\Gamma$, where $c$ is the speed of light and $\Gamma$ is the free-space decay rate of the observed atomic transition. If $R=n\lambda/2$, where $n$ is an integer and $\lambda$ is the wavelength of the transition, an atom located in the center of curvature of the mirror is at a node of the vacuum mode density and inhibition is observed. Conversely, if the radius is $R=n\lambda/2+\lambda/4$, an atom located in the center of curvature is at an anti-node, and enhancement is observed. If the mirror is not a hemisphere ($\text{NA} = 1$), but a spherical mirror with $\text{NA} < 1$ only a partial modification is expected. Fig.~\ref{fig:modification_emission} shows the expected modification of the decay rate of a dipole transition depending on the numerical aperture of a perfectly reflective spherical mirror. Furthermore, in addition to the modification of the spontaneous emission, a hemispherical mirror is predicted to lead to ground and excited state level shifts~\cite{hetet2010qed,hetet2020err}. 

In this article we present the new `Panopticon'~\footnote{The Panopticon setup owes its name to a type of institutional building proposed by the English philosopher and social theorist Jeremy Bentham~\cite{bentham1791panopticon}. In this building, a single watchman located in a central tower can observe all the inmates of the institution, which are located in cells with optical clearance around the tower. Michel Foucault notably used the Panopticon as a metaphor for the modern surveillance state~\cite{fontana2013foucault}. } apparatus for the integration of a high-quality hemispherical mirror ($\text{NA} \approx 1$) and a Paul trap in order to realize the situation described above. In addition to the fundamental components of such a setup, i.e., the hemispherical mirror and a Paul trap, a high numerical aperture lens ($\text{NA} = 0.7$) is used to collimate the light emission and direct it to a detector, as shown in Fig.~\ref{fig:panopticon_scheme}. The design and construction of such an apparatus presents multiple technical challenges which are discussed in the following:
\begin{itemize}
\item In section \ref{sec:mirror} we discuss the construction and characterization of the hemispherical mirror. The effects discussed above vanish rapidly when the mirror deviates from a perfect hemisphere. The target RMS surface error for the fabricated mirror is $\lambda/10$ over the whole mirror surface. Furthermore, we describe how thermal expansion can be used to adjust the radius of the mirror in order to access both inhibition and enhancement of the spontaneous emission.

\item In section \ref{sec:lens} we discuss the design and characterization of the high-NA lens used for light collection. The lens design aims at diffraction-limited performance, enabling interference and imaging experiments such as the ones presented in Refs.~\onlinecite{obvsil2019multipath,araneda2018interference,araneda2018fundamental}.

\item In section \ref{sec:trap} we discuss the design of the Paul trap. The trap must provide full optical clearance between the trapped ion and the hemispherical mirror. Additionally, it must provide optical clearance for the high-NA lens and for the required laser beams. Ideally, the trap design should allow to trap several ions simultaneously in a stable fashion and provide enough flexibility for multi-ion quantum optics experiments such as the one presented in Ref.~\onlinecite{araneda2018interference}.

\item In section \ref{sec:integration} we discuss the integration of the mirror, trap and lens. The position of the optical elements should be adjustable in-situ, with nanometer precision.

\item In section \ref{sec:vacuum_vessel} we present the design of the vacuum vessel. All the above-mentioned elements have to be placed in an ultra high vacuum environment, which contains an atomic source and provides the required electrical connections and optical access.
\end{itemize}

Besides the fundamental interest in measuring effects predicted by quantum electrodynamics~\cite{hetet2010qed,hetet2020err}, being able to control the spontaneous emission of a dipole transition has several applications. For example, reducing the decay rate by two orders of magnitude on a dipole transition would increase the $T_1$ time of the transition up to microseconds, allowing the implementation of a qubit on such a transition. Using a dipole transition would be beneficial because it permits fast qubit control without the use of strong laser beams. Another application can be found in laser cooling, as tuning the decay rate of a dipole transition would allow to directly modify, e.g., the Doppler cooling rate~\cite{castin1989limit,metcalf2003laser}. Furthermore, Doppler cooling typically needs repumper laser beams to depopulate additional states connected to the cooling transition~\cite{metcalf2003laser}. By suppressing the spontaneous decay through these additional channels, it would be possible to Doppler cool without the need of repumping beams.

The system that we present here also provides high light collection efficiency. For an optimally oriented linear dipole, i.e., with its dipole moment oriented orthogonal to the optical axis used for collection, the expected single mode photon collection efficiency is 31.7\,\%~\cite{sattler2016handbook}. Remarkably, given the high quality of the overall optical system, with wavefront errors below $\lambda/10$ for all the components, the collected light can be captured in a single mode and coupled efficiently to a single mode fiber. The high collection efficiency and the cleanness of the optical mode is expected to improve the signal to noise ratio of any optical measurement of atomic properties. Additionally, the improved collection and absorption rates enabled by this setup could be used to implement a quantum network without cavities~\cite{hucul2015modular,stephenson2019high}.

We note that some of the features exhibited by a setup employing a hemispherical mirror, such as high collection efficiency and improved single photon absorption, can be achieved using an atom in the focus of a parabolic mirror~\cite{alber2013qed,maiwald2012collecting}. Indeed, in Ref.~\onlinecite{maiwald2012collecting} collection efficiencies as high as 54.8\,\% have been reported using a parabolic mirror. However, in this approach, the enhancement and inhibition of the spontaneous emission due to QED effects vanish for macroscopic parabolic mirrors~\cite{alber2013qed}. Other approaches consist in using a combination of a high-NA lens collecting the light emitted by an atom and a mirror that retro-reflects the collected light. Although changes in the decay rates have been observed using this method~\cite{eschner2001light}, the modifications are limited by numerical apertures achievable in trapped ion systems for the collection lens. Furthermore, previous attempts to position an ion in the center of a spherical mirror have demonstrated stable trapping and improved light collection~\cite{shu2009trapped}, but they have not achieved the mirror surface quality, radius tunability and numerical aperture required to show inhibition or suppression of the spontaneous emission.

\section{The Panopticon setup}\label{sec:panopticom}

\begin{figure}
\centerline{\includegraphics[width=0.8\columnwidth]{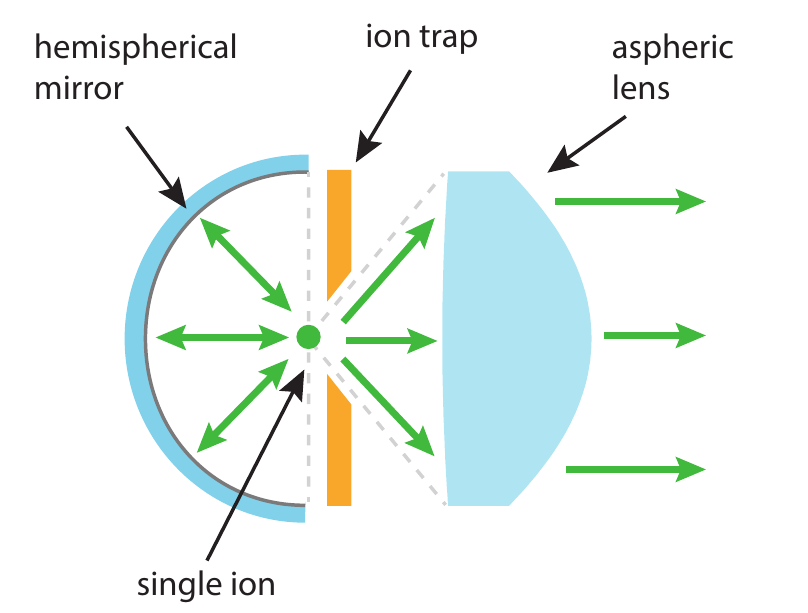}}
\caption[Main components of the Panopticon setup]{\textbf{Main components of the Panopticon setup.} The green arrows show the propagation of the light emitted by the trapped atomic ion.
}
\label{fig:panopticon_scheme}
\end{figure}

The main feature of the Panopticon setup is the ability to confine a single trapped atomic ion in the center of a hemispherical mirror. The ion trap needs to provide full optical access between the trapped ion and the mirror. Additionally, in order to capture a large portion of the emitted field, a high-NA aspheric lens ($\text{NA} = 0.7$) is positioned opposite to the mirror, such that its focal point lies at the position of the ion. Therefore, the trap also needs to provide optical access for the solid angle captured by the asphere. Fig.~\ref{fig:panopticon_scheme} shows a schematic with the main components of the Panopticon setup. In the following sections we describe the design, construction and characterization of each of these main components, as well as their integration and positioning.

\subsection{The hemispherical mirror}\label{sec:mirror}

The main component of the Panopticon setup is the concave hemispherical mirror. Results about the fabrication and characterization of this mirror have been previously reported in Ref.~\onlinecite{higginbottom2018fabrication}. To observe a large modification of the spontaneous emission of an atom located in its center, the surface of the mirror can deviate only minimally from that of a sphere over its full numerical aperture. To our knowledge, the most precise macroscopic spherical objects ever fabricated are spheres with a surface deviation of only 17~nm peak-to-valley~\cite{turneaure1989gravity}. This surface quality is achieved by randomly rotating the sphere between two polishing tools, for periods of several days. Unfortunately, this technique cannot be applied to concave spherical surfaces, and in general, until now, there was no technique able to produce a surface precision similar to that of a convex sphere.

\begin{figure}[ht]
\centering
\includegraphics[width=0.8\columnwidth]{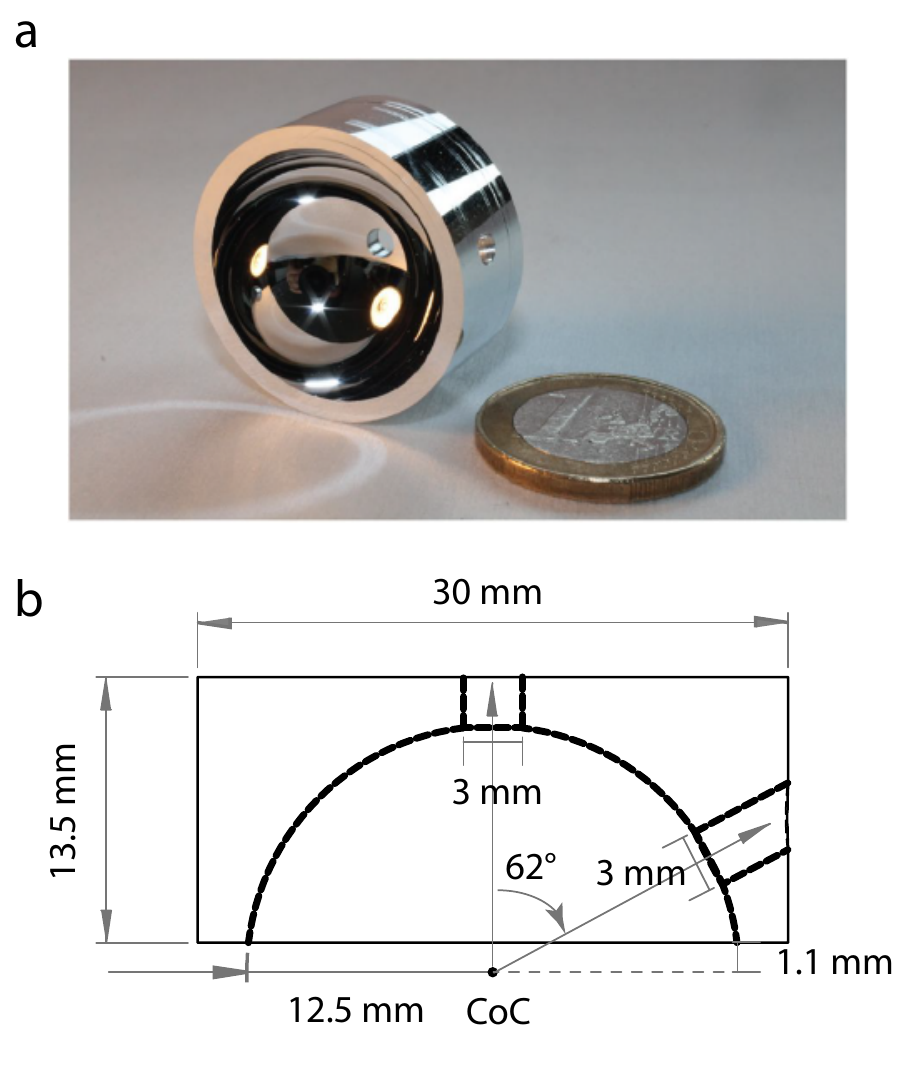}
\caption[Fabricated hemispherical mirror]{\textbf{Fabricated hemispherical mirror.} \textbf{a.} Photograph of a fabricated hemispherical mirror. A euro coin is used for comparison. \textbf{b.} Schematic showing a cross section of the mirror with its relevant dimensions. Note the 3~mm diameter drilled holes which provide laser access to the center of curvature (CoC) of the mirror, where the ion will be located. Reproduced from Higginbottom et al. ``Fabrication of ultrahigh-precision hemispherical mirrors for quantum-optics applications'', Sci Rep 8, 221 (2018); licensed under a Creative Commons Attribution (CC BY) license."
}
\label{fig:photo_mirror}
\end{figure}

In collaboration with the Australian National University (ANU) we have produced hemispherical mirrors with RMS form errors consistently below 25~nm, a maximum peak-to-valley error of 88~nm, and a radius of curvature $\approx\,12.5$~mm. The mirrors were fabricated by diamond turning from a cylindrical aluminium 6061 substrate using a CNC (computer numerical control) nano-lathe \footnote{Nanotech 250UPL, Moore Precision Tools}. To achieve low surface error, an in-situ white-light interferometer was implemented on the lathe, permitting the calibration of the tool with sub-nanometer precision prior to the machining. A detailed description of the fabrication and characterization of the mirrors can be found in Refs.~\onlinecite{higginbottom2018fabrication, DanielThesis}. 

\begin{figure}
\centerline{\includegraphics[width=0.9\columnwidth]{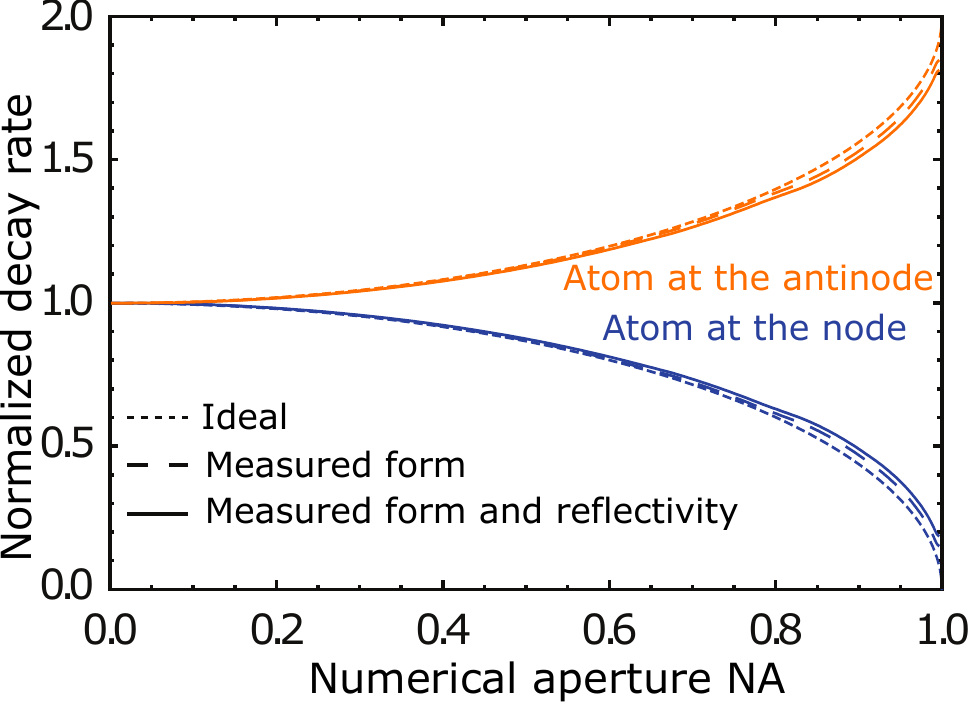}}
\caption[Expected performance of the fabricated mirror]{\textbf{Expected performance of the fabricated mirror.} Modification of the spontaneous emission rate considering the achieved form, and the achieved form and reflectivity of the substrate. The numerical aperture $\text{NA} = 0.996$ is not distinguishable from $\text{NA} = 1.0$ in the plot. Considering form and reflectivity, the maximum expected enhancement and inhibition of the spontaneous emission rate correspond to 88\,\% of the ideal value.
}
\label{fig:exp_mirror}
\end{figure}

Fig.~\ref{fig:photo_mirror} shows one of the fabricated mirrors, with a surface RMS error of 18.1~nm ($\lambda/27$ for $\lambda$ = 493~nm) and a peak-to-valley error of 116.5~nm estimated using stitch interferometry (see Ref.~\onlinecite{higginbottom2018fabrication} for details). Two drilled holes (3~mm diameter) provide laser access to the center of curvature, where the ion will be located (see Section~\ref{sec:full_optical_setup}). One of the holes is located along the optical axis of the mirror and the other at $62^\circ$ with respect to the optical axis. The NA of the mirror is 0.996 (half aperture 85$^\circ$), which is not a limitation of the fabrication process and is chosen such as to provide 1.1~mm optical clearance to the trapped ion. This gives laser access orthogonal to the mirror's optical axis. The reflectivity of the mirror is set by the reflectivity of the substrate (0.92 for aluminium 6061 at 493~nm); it could be improved by applying a highly reflective thin film coating. However, the concave shape of the mirror makes it difficult to realize this in a uniform fashion, and special techniques would be required \footnote{New developments in Atomic Layer Deposition (ALD) have shown promising results in uniform coating of complex 3D structures, making it possible to reach reflectivities close to 100\,\% without affecting the precision of the mirror surface. See for example \url{https://www.laseroptik.de/en/coating-guide/production-methods/ald}}. Fig.~\ref{fig:exp_mirror} shows the expected modification of the spontaneous emission considering the measured surface form, the reflectivity and the effect of the drilled holes. Considering form and reflectivity, the maximum expected enhancement and inhibition of the spontaneous emission rate in the 493~nm transition in $^{138}$Ba$^+$ correspond to 88\,\% of their ideal values. The main deviation from the ideal case comes from the limited reflectivity of the material.

The mirrors have been shown to be resilient to a cycle of vacuum bake-out under typical conditions (maximum temperature of 200 $^\circ$C).

\subsubsection{Temperature stabilization and tuning of the radius of curvature}
\begin{figure*}
\centerline{\includegraphics[width=0.9\textwidth]{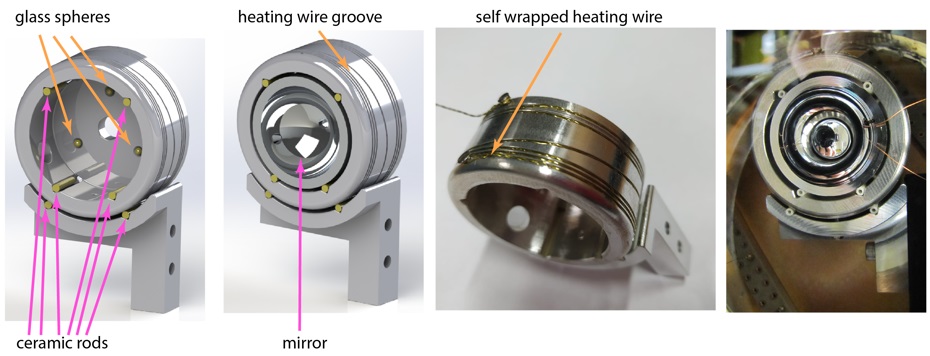}}
\caption{\textbf{Mirror holder and heater.} Components of the mirror's aluminium holder and heater used to achieve a mirror temperature stability below 1 mK. The right-most photograph shows the setup used during the temperature stability tests performed in vacuum, with a thermistor attached directly to one of the mirrors.
}
\label{fig:mirror_holder}
\end{figure*}

To tune the radius of curvature of the mirror, we rely on the uniform thermal expansion of the substrate. The coefficient of thermal expansion of aluminium 6061 at room temperature is $23.5\times 10^{-6}$ K$^{-1}$~\cite{hodgson1990asm}. A change in radius of curvature of $\lambda/4\approx 123.3$ nm (with $\lambda = 493$ nm) requires a temperature modification of 0.42 K. The mirror was heated while measured with a single-shot white-light ZYGO interferometer; this showed radius tunability over the desired range with no measured surface distortion due to thermal expansion. Nevertheless, small variations in environmental temperature in the experiment may lead to drifts of the radius of curvature, therefore, active stabilization of the temperature of the mirror is necessary. 

To control and stabilize the temperature of the mirror we designed a resistively heated holder, shown in Fig.~\ref{fig:mirror_holder}. The holder is heated by applying current to a self-wrapped heating wire \footnote{Insulated nichrome wires, LakeShore Cryotronics$^\text\textregistered$ NC-32}. The total resistance of the heating wire is 65 Ohm, and the current applied varies between 0.1 - 0.2 A depending on the set temperature. The temperature of the holder is measured in the back of the holder using two temperature sensors \footnote{NTC Thermistors BC101B1K, Littelfuse$^\text\textregistered$}, which are glued to the holder \footnote{In these tests a silver-filled epoxy is used, Kurt J. Lesker KL-325K}. The mirror is not in direct contact with the holder, it lies on ceramic rods \footnote{Ceramic tube Allectra 358S-TUBE-20, diameter 2 mm} and glass spheres \footnote{Glass sphere diameter 2 mm} (see Fig.~\ref{fig:mirror_holder}), providing a distance of 1.1 mm between the outer surface of the mirror and the inside of the holder. This reduces the contact thermal conductivity between the mirror and the holder while maintaining the radiative heat transfer, and provides a homogeneous temperature over the volume of the mirror, albeit with a slow temperature tuning rate. The temperature is stabilized with a feedback loop using a high-precision PID \footnote{BelektroniG BTC-LAB-A2000}. In-vacuum tests have shown mirror temperature fluctuations below 1 mK over 10 hours, for different set points between 293 K and 303 K. The heating time constant of the mirror has been measured to be 5.39 hours, while the cooling time constant (cooling is achieved passively) is 3.89 hours. These times may be reduced by increasing the thermal contact between the mirror and the holder, which would, however, lower the temperature stability. The PID can be operated in sample and hold or constant current regime, which is useful for sensitive atomic measurements where a constant magnetic field could be required. The achieved temperature stability corresponds to fluctuations in the mirror's radius of curvature below 20 nm, that are lower than the measured surface error of the mirror. As mentioned above, the mirror is grounded. This is done to avoid any charging that could affect the trapping potential. The grounding is done with a thin (25 $\mu$m diameter) gold wire in the front side of the mirror, fixed with conductive epoxy. The grounding wire has negligible effect in the temperature stability and heating and cooling constants.

\subsection{The aspheric lens}\label{sec:lens}
To capture and collimate the light emitted by the ion, an aspheric lens is located opposite to the hemispherical mirror, as shown in Fig.~\ref{fig:panopticon_scheme}. The main advantages of using an aspheric lens instead of a multi-lens objective are the simple, compact design and the reduced spherical and other optical aberrations. A low-aberration optical system is a crucial requirement for future experiments related to spatial properties of emitted and absorbed photons, for interference experiments with photons emitted by atoms in different traps coupled with optical fibers.

\begin{figure}[ht]
\centering
\includegraphics[width=0.6\columnwidth]{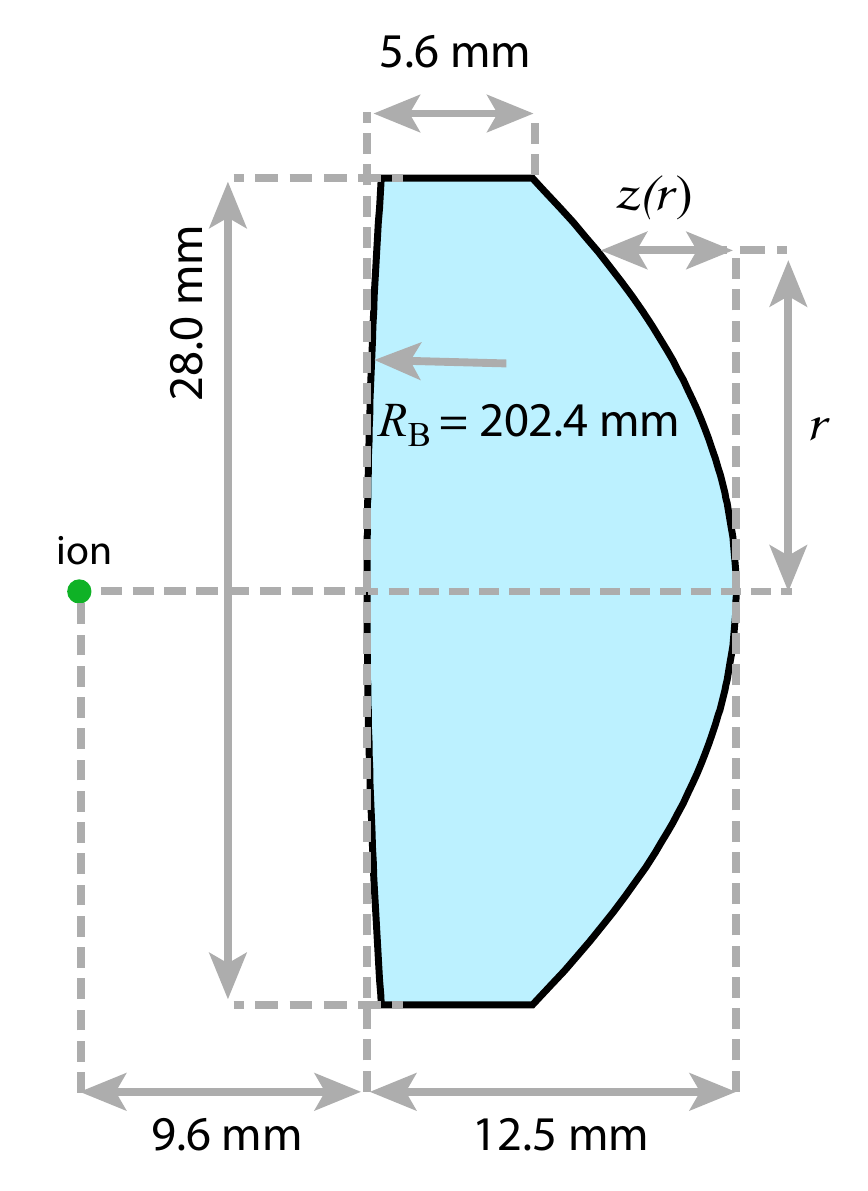}
\caption[Aspheric lens dimensions and parameters]{\textbf{Aspheric lens dimensions.} The surface facing the ion is spherical with a radius of curvature $R_\text{B}$, whereas the aspheric surface is defined by Eq.~\eqref{EQ:asphere}. The ion is positioned 9.6 mm away from the front surface. The aspheric surface design parameters are listed in Table~\ref{TAB:lens_parameters}.
} 
\label{fig:asphere}
\end{figure}

The asphere, designed and fabricated by Asphericon GmbH, has an $\text{NA} = 0.7$, a working distance of 9.60 mm and an effective focal length of 16.05 mm. The design was optimized using Zemax OpticStudio to achieve a diffraction-limited performance with ultra-low wavefront aberrations at 493 nm. The dimensions of the designed asphere are shown in Fig.~\ref{fig:asphere}. The surface facing the ion has a designed constant radius of curvature $R_\text{B} = 202.353$\,mm, whereas the opposite surface is a rotation symmetric asphere defined by the equation
\begin{eqnarray}
z(r) = \frac{r^2}{R_\text{F}\left( 1+ \sqrt{1-(1+k)\frac{r^2}{R_\text{F}^2}} \right)}+\sum\limits_{i=2}^{8} A_{2i}r^{2i}\,.
\label{EQ:asphere}
\end{eqnarray}
The values of parameters $R_{\text{F}}$, $k$ and $A_{2i}$ are listed in Table~\ref{TAB:lens_parameters}.

%%%%%%%%%%%%%%%%%%%%%%%%%%%%%%%%%%%%%%%%%%%%%%%%%
\begin{table}
\centering
\caption{Aspheric lens design parameters }
\begin{ruledtabular}
\begin{tabular}{c c}
Parameter & Value \tabularnewline
\hline 
$R_\text{F}$ & $14.56\,\text{mm}$ \tabularnewline

$k$ & $-0.776$ \tabularnewline

$A_4$ & $3.2022806\times 10^{-6}\,\text{mm}^{-3}$ \tabularnewline
 
$A_6$ & $-2.9002661\times 10^{-8}\,\text{mm}^{-5}$ \tabularnewline

$A_8$ & $-9.6249910\times 10^{-11}\,\text{mm}^{-7} $\tabularnewline

$A_{10}$ & $-1.0236456\times 10^{-13}\,\text{mm}^{-9}$ \tabularnewline

$A_{12}$ & $ 4.5511459\times 10^{-16}\,\text{mm}^{-11}$ \tabularnewline

$A_{14}$ & $ 3.3201252\times 10^{-18}\,\text{mm}^{-13}$ \tabularnewline

$A_{16}$ & $-8.7645298\times 10^{-21} \,\text{mm}^{-15}$ \tabularnewline
\end{tabular}
\end{ruledtabular}
\label{TAB:lens_parameters}
\end{table} 
% YC: I changed the sign of A_4, can you double-check?

\begin{figure}
\centering
\includegraphics[width=0.6\columnwidth]{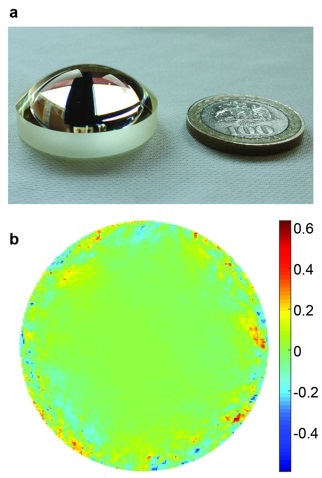}
\caption[Fabricated aspheric lens]{\textbf{Fabricated aspheric lens.} \textbf{a.} Photograph of one of the three fabricated aspheric lenses together with a hundred Chilean pesos coin, giving the scale (the coin has a diameter of 23.25 mm). \textbf{b.} Measured wavefront distortion of one of the aspheres. Numbers in the color scale are in units of $\lambda = 493\,$nm. Provided by Asphericon GmbH.
}
\label{fig:fab_aspheres}
\end{figure}

Three pieces of the designed asphere were fabricated using a S-TIH53 glass substrate \footnote{Ohara Corporation, quoted transmission of 0.951\,\% at 488 nm. Data sheet at \url{https://www.oharacorp.com/pdf/estih53.pdf}}, with refractive index $n_\text{g}\approx 1.87$ ($\lambda = 493$\,nm) and Abbe number $v_g = 23.59$ ($\lambda = 546$\,nm). The manufacturing process includes CNC grinding and polishing as a first step. Thereafter, ion beam figuring is used to reduce surface irregularities~\cite{xie2013ion}. In this process, ion beams are directed at specific regions of the surface where surface errors have been detected. Finally, an anti-reflective coating for 493 nm is applied using plasma-assisted physical vapour deposition~\cite{schneider2000recent}. All these steps were performed by Asphericon GmbH. Fig.~\ref{fig:fab_aspheres}a shows a photograph of one of the aspheres and Fig.~\ref{fig:fab_aspheres}b shows the measured wavefront distortion \footnote{The wavefront distortions were measured using a high-resolution wavefront sensor, Phasics SID4-307, at Asphericon GmbH}. RMS wavefront distortions below 37 nm ($\lambda/14$) and peak-to-valley distortions below 700 nm over the full numerical aperture were achieved for the three aspheres. From Fig.~\ref{fig:fab_aspheres}b it is clear that the largest wavefront distortions occur close to the lens edge. In fact, by reducing the aperture of the lens to $\text{NA} = 0.65$, the RMS wavefront error is reduced to $\lambda/24$. The reflection from each coated surface was measured to be smaller than 0.4\,\% for incidence angles up to $40^\circ$, and the on-axis transmission of the lens was $\approx 95\,\%$, in agreement with the expected values. Both measurements were done with a laser beam with a 493 nm wavelength.

The lens is mounted in an aluminium holder, which allows for laser access to the focal point of the lens from several directions (see Section \ref{sec:full_optical_setup}), and with negligible reduction of the numerical aperture.

\subsection{The ion trap}\label{sec:trap}
\begin{figure}
\centerline{\includegraphics[width=1.1\columnwidth]{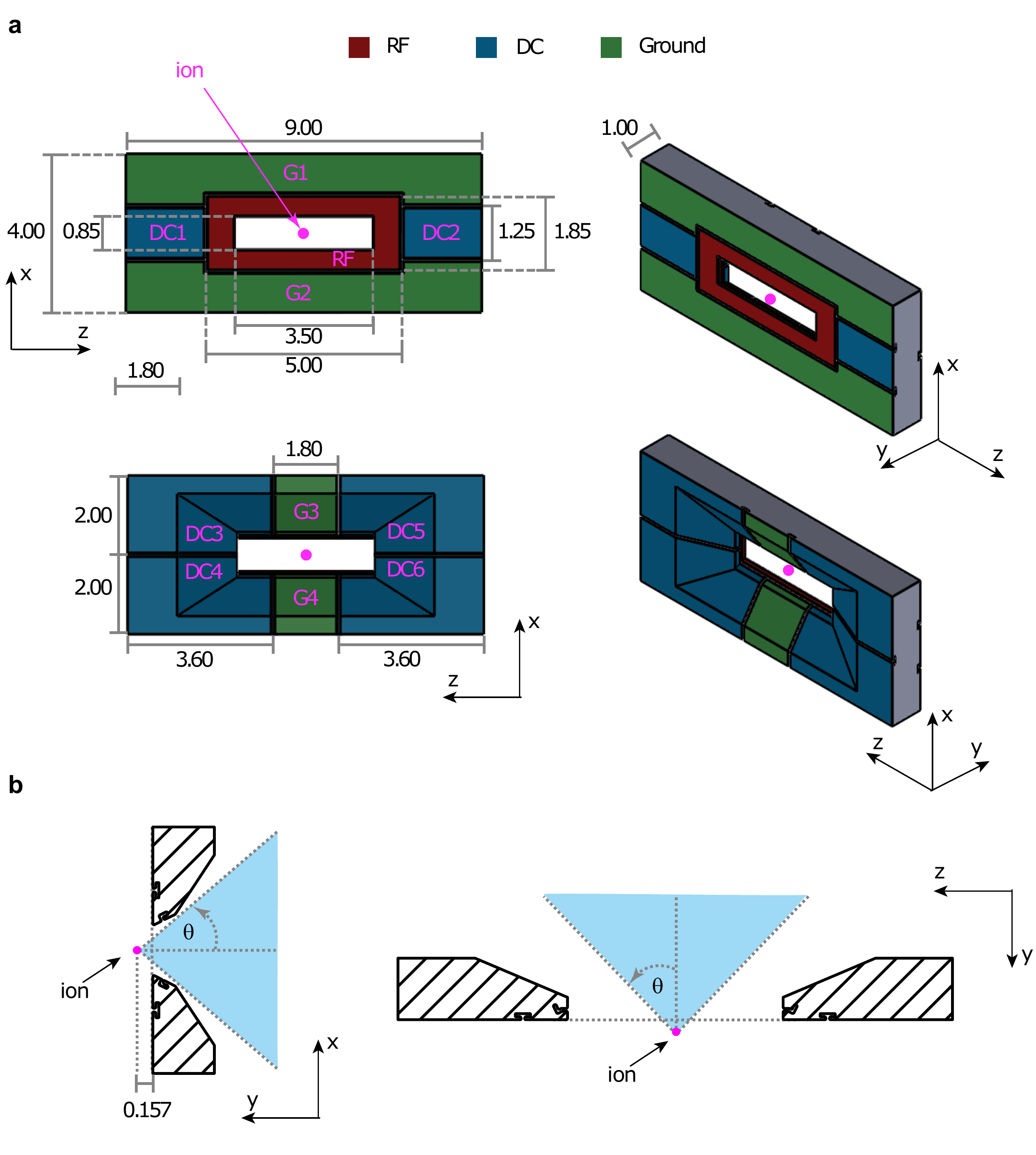}}
\caption[Ion trap design]{\textbf{Ion trap design.} \textbf{a.} Simplified design of the trap. The RF, DC and ground (G) electrodes lie on the facets of a 3D-printed substrate. All dimensions in mm. White text indicates electrode names. \textbf{b.} Transverse cuts of the trap showing the optical clearance. The trap design provides full clearance towards the hemispherical mirror ($y>0$). On the other side ($y<0$), the optical clearance is slightly larger than required by the aspheric lens with $\text{NA}=0.7$, corresponding to a $\theta =\arcsin (\text{NA}) \approx 44.4^\circ$ opening angle (blue area). The RF pseudo-potential minimum, and therefore the position of the ion, is located at a distance $y$ = 0.157 mm from the front plane of the trap.
} 
\label{fig:scheme_fempto_trap}
\end{figure}

\begin{figure}
\centerline{\includegraphics[width=0.6\columnwidth]{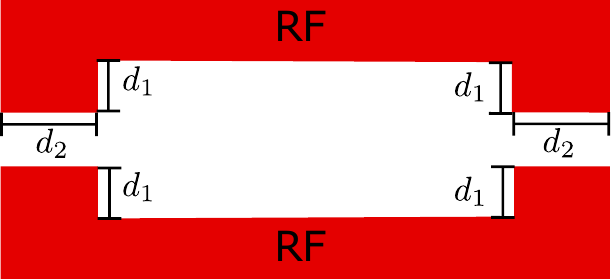}}
\caption[Ion trap design]{\textbf{Slot shape optimization.} The ring shaped RF is the result of optimization starting from two separated $\pi$-shaped RF electrodes, and varying the lengths $d_1$ and $d_2$ while keeping the other dimensions of the electrode fixed. 
} 
\label{fig:optim}
\end{figure}

The main challenge in the design of an ion trap compatible with the Panopticon setup is that the trap should provide full optical access not only between the trapped ion and the hemispherical mirror, but also between the ion and the aperture defined by the aspheric lens. This could be achieved using, for example, a needle trap~\cite{siverns2017ion}. However, this approach allows the trapping of only one ion without excess of micromotion and displacement of the trapped ions. It is desirable to be able to trap several ions without micromotion, since the setup could then be used for a broader range of quantum optics experiments. Some comply partially with the criteria for optical access and stable trapping of several ions, including, for example, the miniaturized segmented ``High Optical Access Trap 2.0'' developed by Sandia National Laboratories \footnote{Information on Sandia National Laboratories' High-Optical-Access trap can be found at \url{https://www.osti.gov/servlets/purl/1239095}}. However, this trap only possesses a clear aperture equivalent to $\text{NA} = 0.25$. The approach that we take here is the fabrication of a monolithic slotted pseudo-planar macroscopic trap. A simplified scheme of the trap geometry, with only the minimal electrode configuration needed to trap an ion, is shown in Fig.~\ref{fig:scheme_fempto_trap}. Later in this article, we discuss the actual fabricated trap with additional features. In the design, an ion is trapped above the front surface, providing full clearance on the side of the hemispherical mirror and, through the slot, clearance corresponding to the numerical aperture $\text{NA} = 0.7$ of the aspheric lens. The geometry shown in Fig.~\ref{fig:scheme_fempto_trap} generates a confining RF pseudo-potential in the $x$ and $y$ directions, whereas trapping along the $z$ direction is provided by the DC potential generated primarily by the DC electrodes DC1 and DC2. The DC voltages applied to the electrodes DC3, DC4, DC5 and DC6 contribute weakly to the trapping potential in the $z$ direction and are meant for micromotion compensation. The ring shape of the RF electrode, although counter intuitive, is the result of systematic optimization, starting from two $\pi$-shaped separated RF electrodes producing a cross-shaped slot (see Fig.~\ref{fig:optim} ). Varying the dimensions of the slot and keeping a fixed slot vertical (along $x$) aperture. The obtained shape with a rectangular slot exhibits a reduced residual axial ($z$ direction) pseudo-potential, allowing chains of $\sim$ 20 ions to be trapped with negligible axial micromotion.

\begin{figure}
\centerline{\includegraphics[width=1.0\columnwidth]{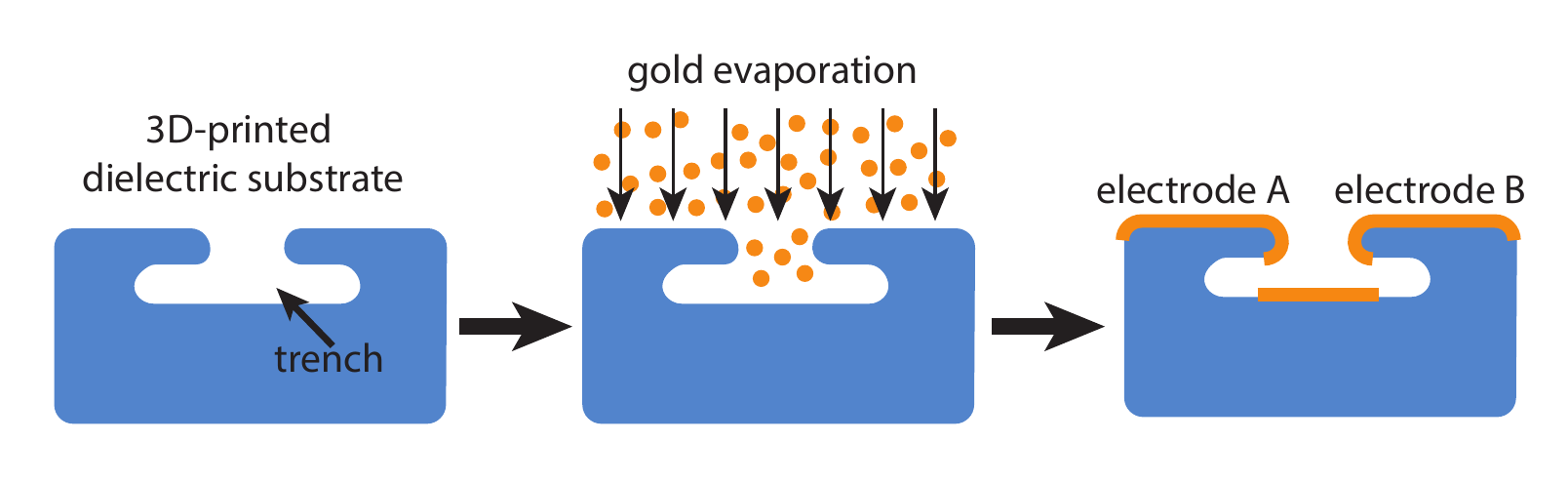}}
\caption[Electrode fabrication through 3D-printing and coating]{\textbf{Electrode fabrication through 3D-printing and coating.} First, a dielectric substrate is ``3D-printed" to create the trenches needed to separate electrodes. Then, the surfaces are coated with a conductor (gold) in an evaporation step. The profile of the trenches combined with the directionality of the evaporation step provides electric insulation between different regions, defining trap electrodes. The recess in the profile allows to apply evaporation from different angles without shorting the electrodes. The dimensions of the trench profile chosen for the fabricated trap are shown in Fig.~\ref{fig:actual_trap}d.
} 
\label{fig:3D_print}
\end{figure}

The fabrication is done with subtractive 3D laser micro-machining technology. The basic idea is to laser-machine a monolithic dielectric substrate to the required shape, including ``trenches'' separating different regions. In the following, we will refer to this technique as `3D-printing'. The surfaces are then coated with gold, creating an independent electrode in each region surrounded by trenches. Fig.~\ref{fig:3D_print} depicts how 3D-printing and coating are combined to create electrodes without the need for shadow masking and reduce the number of angled metal evaporation steps needed. The bottom of the carved trenches can be grounded or connected to one of the contiguous electrodes by shaping ``ramps'' between the bottom and the surface in the 3D-printing design.

The required precision of the dielectric substrate fabrication can be achieved by ``laser carving'' techniques provided by companies such as FEMTOprint S.A. in Switzerland or Translume Inc. in USA. The 3D-printing technology used by these companies employs strongly-focused high-energy laser pulses to locally change the physical structure of the substrate. A chemical process is then used to remove the treated material. Using this technique on, e.g., fused silica, the carved features can reach an average surface roughness below 100 nm, whereas the untouched surfaces can reach an average roughness below 5 nm. Additional polishing can reduce the roughness below 50 nm, though with detrimental effects in surrounding areas. The scale of the surface roughness achieved has been demonstrated to be adequate for use in ion traps with minimum ion-electrode distance of $\sim$ 100 $\mu$m, see for example Ref.~\onlinecite{ragg2019segmented}, whereas in our trap that distance, as described below, is 453 $\mu$m.  A more detailed description of this fabrication technology applied to the fabrication of ions traps can be found in  Ref.~\onlinecite{ragg2019segmented}.

The conductive coating is applied in our clean room by means of electron beam evaporation of a thin layer of titanium ($\sim$ 2 nm), used as an adhesion layer, followed by a thick layer of gold ($\sim$ 200 nm). Gold evaporation is applied from three different angles on each side of the substrate. Gold electroplating could also be used afterwards to increase the thickness of the electrodes to $\sim 5\,\mu$m.

\subsubsection{Simulations of the trapping potential}\label{sec:sim}
The trap geometry presented in Fig.~\ref{fig:scheme_fempto_trap} is the result of a systematic optimization. The different dimensions and positions were varied to maximize the trapping frequencies for a $^{138}\text{Ba}^+$ ion, as well as the trap depth, and to reduce the trap capacitance and the residual axial RF field. The ion-electrode distance, which is constrained by the required optical clearance and laser access to the ion, was maximized. The electric potentials produced by the trap electrodes were simulated via finite element analysis using the software COMSOL Multiphysics 4.4. In these simulations the grounded hemispherical mirror was included.

\begin{figure}
\centerline{\includegraphics[width=1.0\columnwidth]{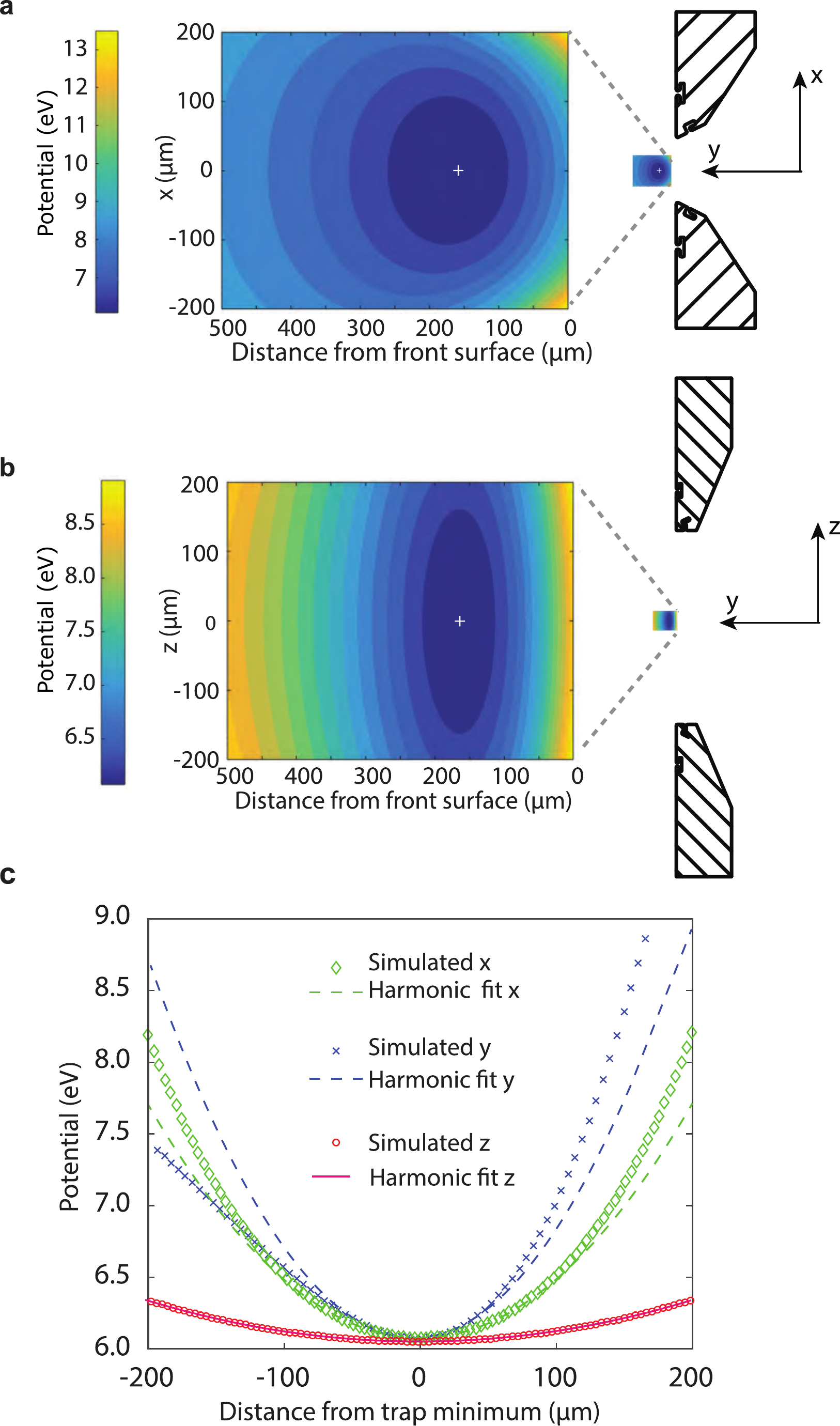}}
\caption[Simulated trapping potential.]{\textbf{Simulated trapping potential.} The parameters are those of `config. 1' in Table~\ref{TAB:femto_sim}. \textbf{a.} Cross section through the $x$-$y$ plane. The white cross shows the position of the minimum. \textbf{b.} Cross section through the $y$-$z$ plane. \textbf{c.} Trapping potential along the $x$, $y$ and $z$ axis. Close to the center, in a range of $\pm 50\,\mu$m, the potentials are well approximated by harmonic potentials.} 
\label{fig:femto_potential}
\end{figure}

Fig.~\ref{fig:femto_potential} shows the total trapping potential. The RF pseudo-potential minimum is located 157 $\mu$m away from the front plane of the trap, and the distance of a trapped ion to the closest electrode is 453 $\mu$m. For comparison, in our current ``Innsbruck-style" blade trap~\cite{RotterThesis} this distance is 707 $\mu$m. Small ion-electrode distances tend to increase the ion heating rate due to, e.g., surface noise~\cite{brownnutt2015ion}. Keeping this distance relatively large is therefore important.

Fig.~\ref{fig:femto_potential}c shows the trapping potentials along each coordinate axis. The potential is symmetric about the trap center along directions $x$ and $z$, but not along $y$, as typically observed in surface traps~\cite{siverns2017ion}. The trapping frequencies can be varied with the voltages applied to the different electrodes. Table~\ref{TAB:femto_sim} shows the simulated trapping frequencies and trap depths obtained for a $^{138}\text{Ba}^+$ ion. The trap depths are in the eV regime, comparable with 3D rod or blade traps~\cite{bruzewicz2019trapped}, which makes the trap suitable for operation at room temperature. 

Simulations show that independent control of the voltages applied to the electrodes DC3, DC4, DC5 and DC6 is sufficient for compensation of micromotion in all directions. Further simulations with different trench geometries and dimensions were carried out. The most relevant trench parameter, i.e., the electrode-electrode separation provided by the trench, was varied between 50 $\mu$m and 150 $\mu$m. Variations in this range have negligible effect on the trapping potential, so that the final dimensions were selected based on breakdown test results described in the following section.

\begin{table}
\begin{center}
\caption[Trap voltages and frequencies]{Trap driving parameters and resulting trap frequencies and depths, obtained by finite elements simulations for $^{138}$Ba$^+$. Three different configurations are shown. DC1 and DC2 refer to the `endcap' electrodes shown in Fig.~\ref{fig:scheme_fempto_trap}, while DC3, DC4, DC5 and DC6 are the electrodes in the back plane. In these three configurations the electrodes G1, G2, G3, G4 and the mirror are grounded.}
\begin{ruledtabular}
\begin{tabular}{l c c c}
 &\textbf{ config.1} & \textbf{config. 2}& \textbf{config. 3}
\tabularnewline
RF freq. $\Omega_{\text{RF}}/2\pi$ (MHz) &  16.0 & 16.0  & 16.0 \tabularnewline
\hline 
RF amplitude $U_{\text{RF}}$(V) &  1000  & 1500 & 2000  \tabularnewline
\hline 
DC1,2 (V) &  200 & 300 & 400 \tabularnewline
\hline 
DC3,4,5,6 (V) & 82  & 123 & 164\tabularnewline
\hline 
$\omega_{x}/2\pi$ (MHz) & 1.33 & 2.16 & 2.96\tabularnewline
\hline
$\omega_{y}/2\pi$ (MHz) & 1.57 & 2.34 & 3.07\tabularnewline
\hline
$\omega_{z}/2\pi$ (MHz) & 0.51 & 0.62 & 0.72\tabularnewline
\hline
Trap depth (eV)         & 2.4  & 4.9  & 8.2 \tabularnewline
\label{TAB:femto_sim}
\end{tabular}
\end{ruledtabular}
\end{center}
\end{table} 

\begin{figure*}
\centerline{\includegraphics[width=1.0\textwidth]{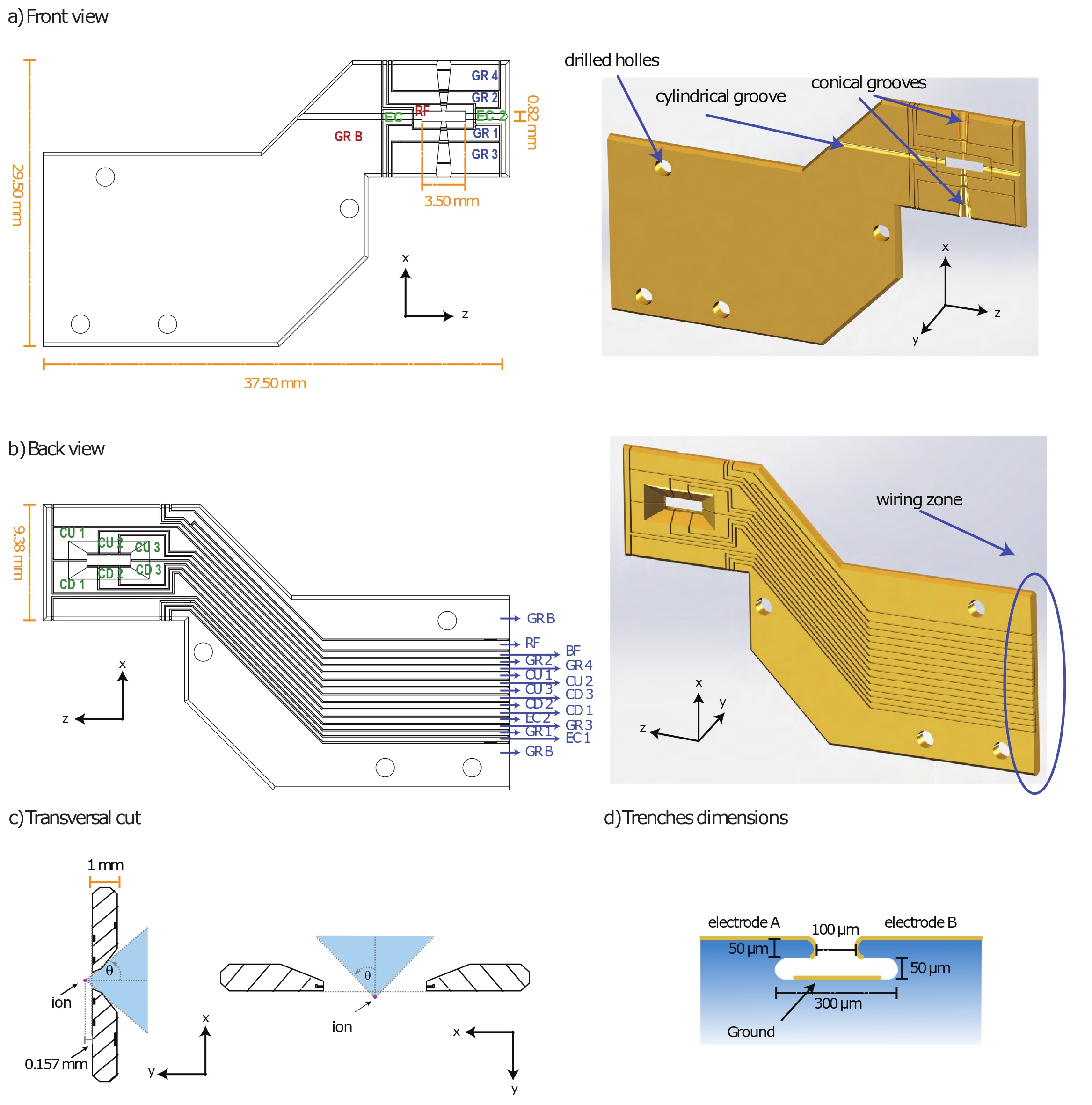}}
\caption[Full design of the ion trap]{\textbf{Full design of the ion trap.} \textbf{a.} Front side of the final trap design. Cylindrical and conical grooves are carved in the substrate in order to minimize the scatter of laser beams propagating close to the surface. All front electrodes extend to the back plane via the edges of the trap. \textbf{b.} In the back plane all the electrodes are extended to one extremity of the substrate, where wire-bonding is used for connection to voltage supplies. \textbf{c.} Transverse cut at the trap center showing the optical clearance provided by the trap. \textbf{d.} Dimensions of the trenches separating the electrodes.
} 
\label{fig:actual_trap}
\end{figure*}

\subsubsection{The fabricated trap}
While the fabricated trap has the same core geometry as presented in Fig.~\ref{fig:scheme_fempto_trap}, it includes additional features. Fig.~\ref{fig:actual_trap} shows schematics and renders of the actual design. The ground electrodes G1 and G2 in the front plane are divided into smaller electrodes GR1, GR2, GR3 and GR4 (Fig.~\ref{fig:actual_trap}a) to add control of the RF potential minimum position while keeping micromotion compensated, allowing the possibility of moving the ion in any direction several micrometers. These electrodes, together with all other electrodes in the front plane (EC1, EC2, RF and GRB), are extended to reach the back plane. There, together with the back plane electrodes CU1, CU2, CU3, CD1, CD2 and CD3, they are prolonged into conductive traces extending beyond the radii of the hemispherical mirror and of the aspheric lens (Fig.~\ref{fig:actual_trap}b). This flat and elongated design provides enough space for wire-bonding, which connects the electrodes to the voltage supplies through a printed circuit board (PCB, see Section~\ref{sec:PCB}). In this way, the optical clearance to the trapping region is not affected by the bonding wires. The extended area is also used to mount the trap to an aluminum holder using through-holes.

Additionally, grooves in the front face are carved in the substrate (during the 3D-printing process). These grooves improve the clearance of the laser beams that propagate close to the surface, reducing the light scattered by the trap. This is of particular importance when addressing single ions using a strongly focused 1.7 $\mu$m laser. The conical vertical grooves shown in Fig.~\ref{fig:actual_trap}a match the divergence of such a beam, with additional 100 $\mu$m clearance from the point where the beam intensity has decayed by a factor of $1/e^2$. The horizontal cylindrical grooves give enough optical clearance for lightly focused axial cooling and optical pumping beams. Fig.~\ref{fig:actual_trap}c shows transverse cuts at the trap center, showing the position of the ion and the optical clearance provided by the design.
The dimension of the trenches separating adjacent electrodes, shown in Fig.~\ref{fig:actual_trap}d, were chosen after performing in-vacuum DC and RF breakdown tests on a simplified trap. The separation between electrodes varied between 50 and 150 $\mu$m. These tests did not shown any sign of electric breakdown between electrodes with DC voltages as high as 700 V and RF voltages as high as 1200~V at 20 MHz. The separation was finally chosen to be 100 $\mu$m. The bottom of the trenches is connected to the common ground GR B. A complete 3D drawing of the designed trap can be found in the supplementary material. Fig.~\ref{fig:trap_photo} shows a photograph of one of the traps fabricated by FEMTOprint S.A. before metalization.

We have performed simulations of the generated trapping potential for the fabricated trap design, including all the aforementioned features. The differences with the results of the simulations of the simplified version presented in section \ref{sec:sim} are negligible.

\begin{figure}
\centering
\includegraphics[width=0.8\columnwidth]{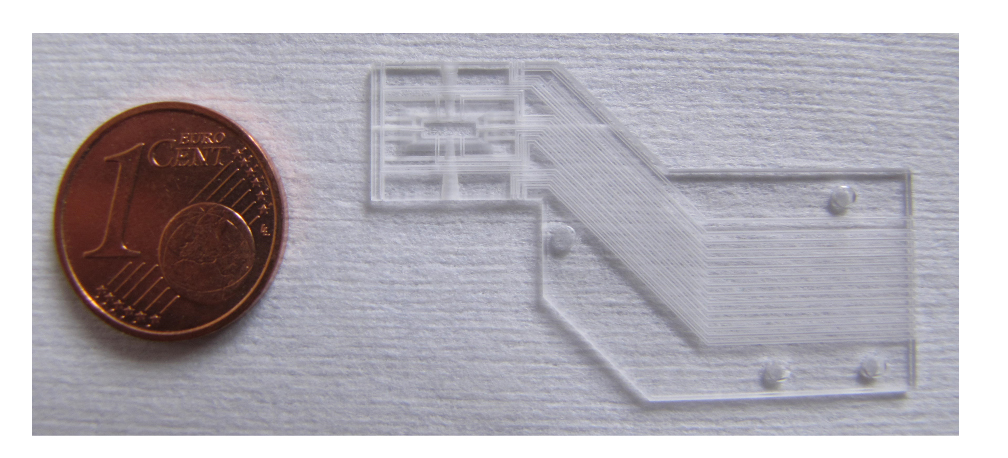}
\caption{\textbf{Fabricated ion trap.} Fabricated ion trap before metalization.
} 
\label{fig:trap_photo}
\end{figure}

\subsubsection{Trap holder, in-vacuum low-pass filter and voltage driving}
\label{sec:PCB}

\begin{figure}
\centerline{\includegraphics[width=0.8\columnwidth]{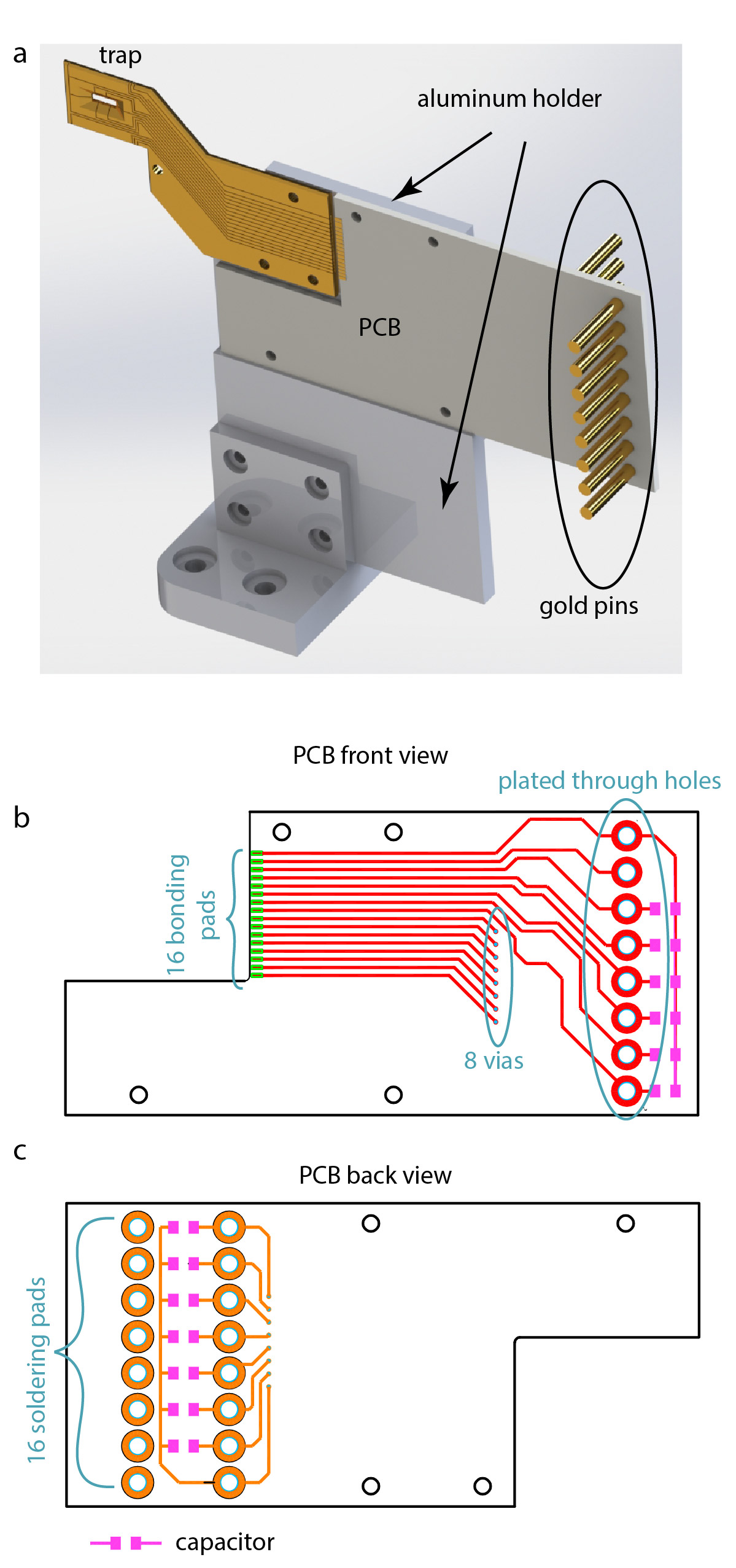}}
\caption{\textbf{Trap holder and filter PCB}. \textbf{a. } Both the trap and the PCB filter are mounted on an aluminum holder. The trap is wire-bonded to the PCB which has conducting paths, capacitors and pins for connection to the DC and RF sources. \textbf{b and c. } Details of the PCB paths, position of the capacitors and pins.
}
\label{fig:trap_holder}
\end{figure}

\begin{figure*}
\centerline{\includegraphics[width=0.7\textwidth]{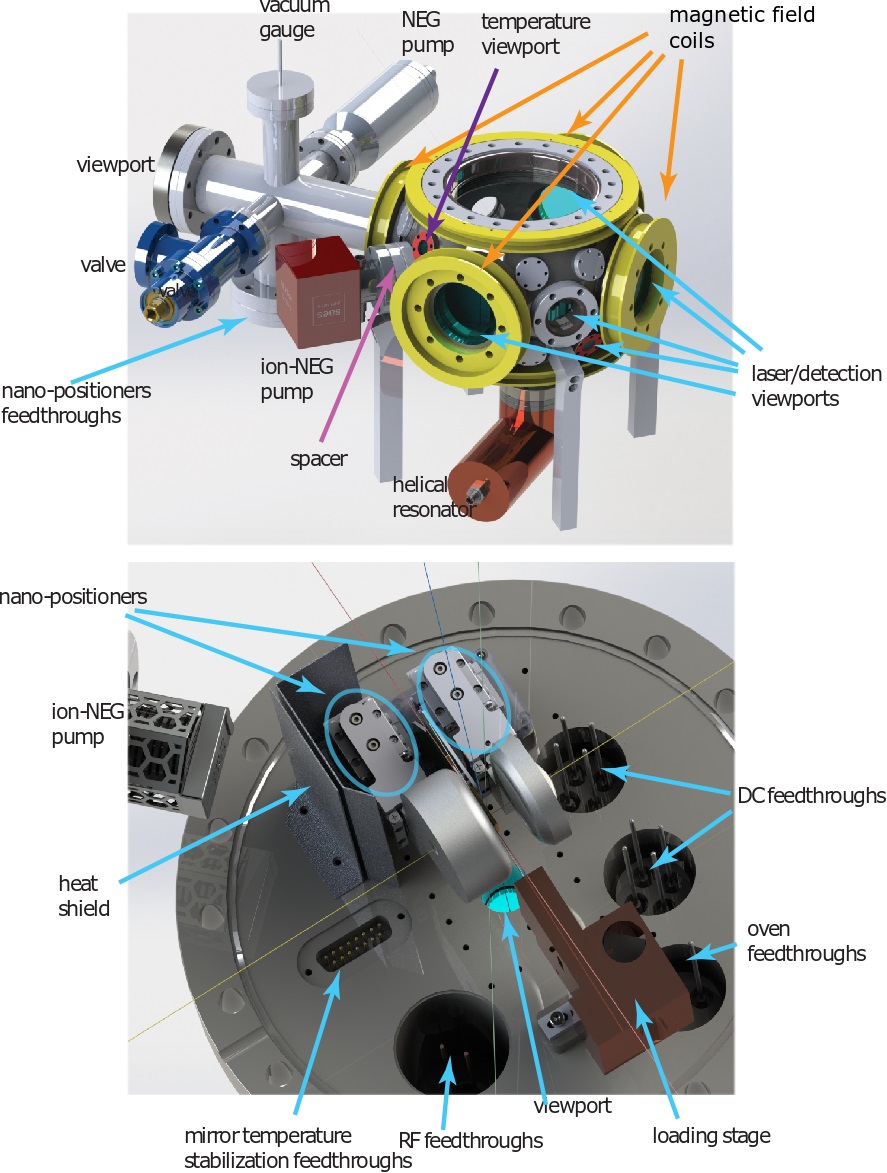}}
\caption[Main components of the vacuum vessel]{\textbf{Main components of the vacuum vessel.} See details in the main text.
}
\label{fig:vacuum_vessel}
\end{figure*}

The ion trap is mounted on an aluminum holder plate (2 mm thickness) using three screws, as shown in Fig.~\ref{fig:trap_holder}a. The holder plate also acts as a heat sink for the trap and is contacted to the body of the vacuum chamber. This plate also holds a filter PCB. The trap is wire-bonded to the PCB using gold wires. The PCB, made of alumina \footnote{Al$_2$O$_3$ 96\,\%, fabricated by ELCERAM a.s., Czech Republic} with a thickness of 1\,mm, has 16 silk-printed gold traces, wire-bond pads, paths and soldering pads \footnote{Screen printable gold conductor composition DuPont 5744R} for DC and RF routing. There are 8 vias to route some of the paths to the opposite side of the PCB. All the DC paths are capacitively coupled to a common ground using 1\,nF high voltage surface mounted capacitors \footnote{Knowles Syfer 0805Y1K00102KST, maximum voltage 1\,kV}, suppressing RF pick up. Additionally, 1\,k$\Omega$ ex-vacuum resistors are used in order to filter high frequency noise induced in the DC electrodes (cutoff frequency 1 kHz). Fig.~\ref{fig:trap_holder}b and c show a schematic view of the PCB. At the edge opposite to the trap, the PCB has 16 gold pins \footnote{Accu-Glass male gold pins type T-2, part number 110008} soldered \footnote{Lead-free solid wire solder, Harris SB61/2POP, 96\,\% Tin, 4\,\% Silver, 430 F}, allowing easy connection and disconnection of each line to copper wires attached to the electrical feedthroughs on the base vacuum flange (Fig.~\ref{fig:vacuum_vessel}). None of the PCB traces or the wires are exposed to the atomic flux from the atom source (see section \ref{sec:atom_source}).

The DC voltages are generated outside the vacuum chamber using a high-precision high-voltage source \footnote{ISEG EHS 82 20p}. The RF voltage is generated using a high-precision signal generator \footnote{Rohde \& Schwarz SMC100A, option B101} followed by a 28 dB amplifier \footnote{MiniCircuits ZHL-1-2W+} and a helical resonator~\cite{macalpine1959coaxial}.

\subsection{System integration}\label{sec:integration}
\label{sec:full_optical_setup}
\begin{figure}
\centerline{\includegraphics[width=1.0\columnwidth]{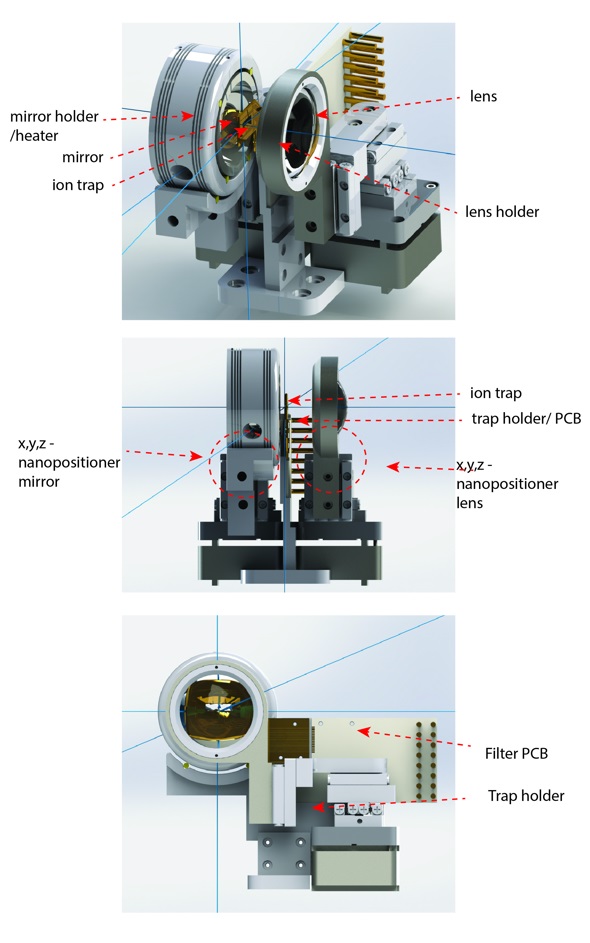}}
\caption[Panopticon optical setup]{\textbf{Panopticon optical setup.} The mirror and lens are mounted on independent $xyz$--nanopositioners, whereas the position of the ion trap is fixed. The blue lines show the four laser axes.}
\label{fig:optical_setup}
\end{figure}

The optical setup, consisting of the hemispherical mirror, the aspheric lens and the ion trap, has to be set in place in a robust and stable manner, while still providing enough degrees of freedom for correct alignment. The holders and positioners have to be compatible with the ultra-high vacuum environment needed to trap single atomic ions in a stable way. To do so, we designed the mounting system shown in Fig.~\ref{fig:optical_setup}. In this setup, all the elements are carefully designed in order to provide the optical clearance required by the mirror and the lens, and laser access. The positions of the mirror and the lens can be set independently using $xyz$-nanopositioners \footnote{SmarAct SLC-1720-S-UHVT} shown in the figure. The positioners have a step size of 1 nm in each direction, a position readout resolution of 1 nm and a maximum displacement of 12 mm.

%\clearpage
%\clearpage
\twocolumngrid

\section{The vacuum vessel}\label{sec:vacuum_vessel}
The Panopticon optical setup is placed in an ultra-high vacuum environment. The vacuum vessel, shown in Fig.~\ref{fig:vacuum_vessel}, is build around an 8''-CF spherical octagon chamber \footnote{Kimball Physics  MCF800-SphSq-G2E4C4A16.}. 

A customized 6-way cross is attached to the main chamber, providing enough flanges to connect a vacuum valve \footnote{VAT 54132-GE02}, a vacuum gauge \footnote{Bayard-Alpert ion gauge, Agilent UHV-24}, the electric 
feedthroughs for the wiring of the nanopositioners, a non-evaporable getter (NEG) pump \footnote{SAES CapaciTorr Z400} and a viewport. The main pumping is done with a combined ion and NEG pump \footnote{SAES NEXTorr D 100-5} attached to the main chamber. During the NEG activation, the pump can reach temperatures close to 450$^\circ\,\text{C}$. This temperature is not compatible with the maximum temperature to which the nanopositioners can be exposed (150$^\circ\,\text{C}$). In order to avoid damaging the nanopositioners, the pump is retracted from the main chamber using a spacer, and a two-layer aluminium heat shield is placed between the pump and the positioners (see Fig.~\ref{fig:vacuum_vessel}). According to finite-element simulations performed by the pump manufacturer this is sufficient to prevent damage of the nanopositioners. We plan to perform additional test to confirm this predictions, and if necessary, the shielding and the pump position will be modified.

The vacuum vessel has all the necessary viewports to provide laser access to the center of the trap using the planned beam directions (see Fig.~\ref{fig:optical_setup}), including a CF-160 viewport on the top of the chamber. All the viewports \footnote{All viewports were fabricated by VACOM GmbH and anti-reflection coated by Laseroptik GmbH} are anti-reflection coated for all the wavelengths needed to load, cool and control $^{138}\text{Ba}^+$ ions, i.e, 413, 493, 614, 650 and 1762 nm. For these wavelengths the reflectivity is below 1\,\%. The viewport used for transmitting the light emitted by the ion and collimated by the aspheric lens has an optical quality surface, with wavefront aberrations below $\lambda/10$ over all the surface (with $\lambda=493$ nm) \footnote{VACOM VPCF63DUVQ-L-LAMBDA10}. An additional Germanium viewport \footnote{VACOM VPCF16GE-K} is attached to the main chamber, to allow monitoring of the temperature of the atom oven using a thermal camera outside the chamber (see Section~\ref{sec:atom_source}).

The electrical connections needed to drive the ion trap, to heat an atom dispenser oven and to measure and control the temperature of the mirror are routed through feedthroughs in the customized CF-160 bottom flange, assembled by Vacom GmbH (Fig. \ref{fig:vacuum_vessel}). This flange has a SUB-15 feedthrough for all the mirror temperature stabilization connections, a two-wire (12 kV, 13 A)  feedthrough for the oven connections, two four-wire (12 kV, 13 A) for the trap DC connections and a 2-wire (5 kV, 25 A) for the RF trap connections. This customized flange also has a CF-16 viewport intended for 1.7 $\mu$m laser addressing of individual trapped ions. The connections for the nano-positioners are in one of the arms of the cross.

\subsection{Atom source}
\label{sec:atom_source}
To provide a source of atoms inside the vacuum chamber, we have designed a loading stage that simultaneously contains a resistively heated Ba dispenser oven and a laser ablation target. The resistively heated oven is a reliable way to produce a flux of neutral atoms in the center of the trap, which can then be ionized using a 413 nm laser through a two-step excitation process\cite{RotterThesis}. This process is, however, slow, and produces an excess of heat inside the vacuum chamber that can have detrimental effects on the operation of ion traps, such as undesired thermal expansion of the electrodes and thermal expansion of the hemispherical mirror.

Laser ablation from a target is an alternative method which has been proved to be more efficient and less detrimental to ion trap operation (see for example Ref.~\onlinecite{leibrandt2007laser}). In this approach, short and strong laser pulses are applied on a Ba target to produce neutral and ionized atoms, which eventually reach the center of the trap. It has been experimentally shown that instead of using a pure Ba target, targets containing BaTiO$_3$ or BaO produce higher Ba$^+$ yields~\cite{olmschenk2017laser}, making loading more efficient while lifting the requirement of an additional photo-ionization laser. At the same time, using BaTiO$_3$ instead of pure Ba prevents fast oxidation of the sample during preparation of the vacuum setup. To guarantee isotope-selective loading, we still plan to use photo-ionization of ablated neutral atoms.

\begin{figure}
\centerline{\includegraphics[width=1.0\columnwidth]{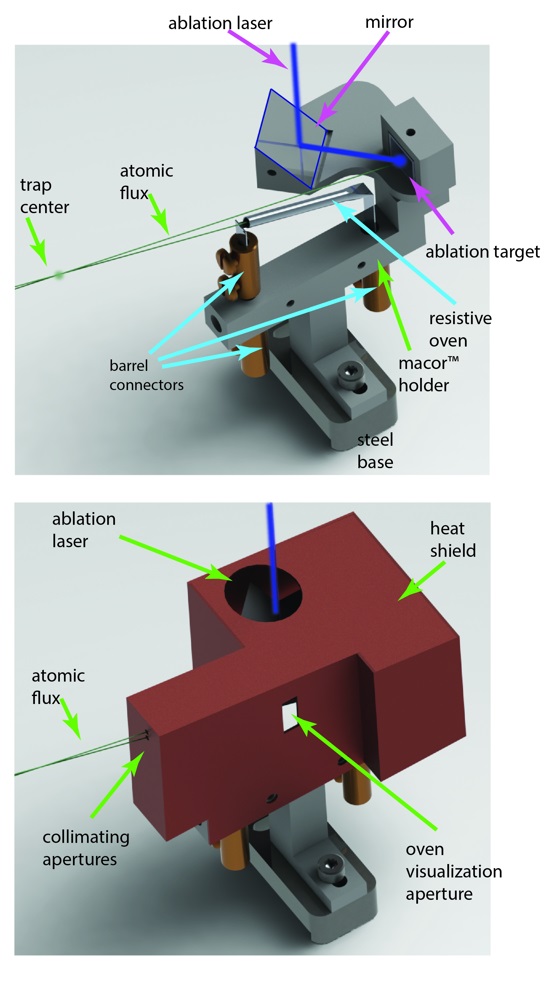}}
\caption[Loading stage]{\textbf{Loading stage.} Design of a loading stage combining a resistive Barium dispenser oven and a target for laser ablation. See details in the main text.
}
\label{fig:loading_stage}
\end{figure}

Fig.~\ref{fig:loading_stage} shows the designed loading stage, where both a resistive dispenser oven \footnote{Custom design fabricated by AlfaVakuo e.U., Austria} and a BaTiO$_3$ ablation target \footnote{Custom design fabricated by Testbourne Ltd, UK} are located. The holder is made of macor$^\text\textregistered$ ceramic, which exhibits a low thermal and electric conductance. Both the target and the oven are enclosed in a copper shield, which simultaneously acts as a thermal radiation shield and prevents the ablated Barium from coating surfaces in the chamber and in the trap. Two circular apertures in the front with 1 mm diameter, one for each atom flux production process, are used to collimate the atomic flux towards the center of the ion trap. As the ablation process spreads atoms in every direction, a mirror and a hole in the shield are used (Fig.~\ref{fig:loading_stage}). This configuration prevents a high atomic flux from exiting the shield and directly coating the top CF-160 viewport used for the ablation laser beam. Although the mirror used for the ablation laser is directly exposed to the atomic flux produced by the ablation producing a systematic degradation on the mirror's reflectivity, this can be compensated directly by increasing the power of ablation the laser. Additionally, there is a small square aperture in the side of the stage, which allows for monitoring of the oven temperature with a thermal camera through an infra-red transmissive Germanium viewport. This can be used to characterize the heating dynamics of the oven in ultra-high vacuum and to optimize the loading process avoiding excess heating~\cite{ballance2018short}.

A broad range of pulsed laser sources can produce the pulses needed for ablation. An example of such a source is a pulsed diode pumped solid state laser (DPSS), with a wavelength of 515 nm and energy per pulse of~170 $\mu$J \footnote{Coherent Flare NX}. This wavelength is compatible with our coated viewports.

\subsection{Vacuum preparation and ex-vacuum elements}
The cleaning of all the vacuum components is done following the procedure described in Ref.~\onlinecite{obvsil2019room}. The main components of the vacuum chamber are pre-baked at 450 $^\circ$C, whereas when adding the the in-vacuum optical system and the nanopositioners, the sytem must be baked at 150 $^\circ$C.

Three pairs of magnetic field coils are attached to the outside of the main chamber flanges in order to provide a homogeneous magnetic field at the center of the trap. These coils may be replaced by rings of permanent magnets in order to achieve a better suppression of RF magnetic field noise. Four legs are attached to the exterior of the vacuum chamber to support it on an optical table.

\section{Summary and outlook}
In this article we have presented the design and construction of a new setup which will allow us to study quantum electrodynamics effects. Furthermore, the setup features an unprecedented single-mode collection efficiency. We have presented the design and construction of the main optical components of the setup, namely a hemispherical mirror and an aspheric lens, and shown how the strict requirements on their optical performance can be fulfilled. Both the mirror and the lens achieve a wavefront distortion below $\lambda/10$. 

\begin{figure}
\centerline{\includegraphics[width=0.7\columnwidth]{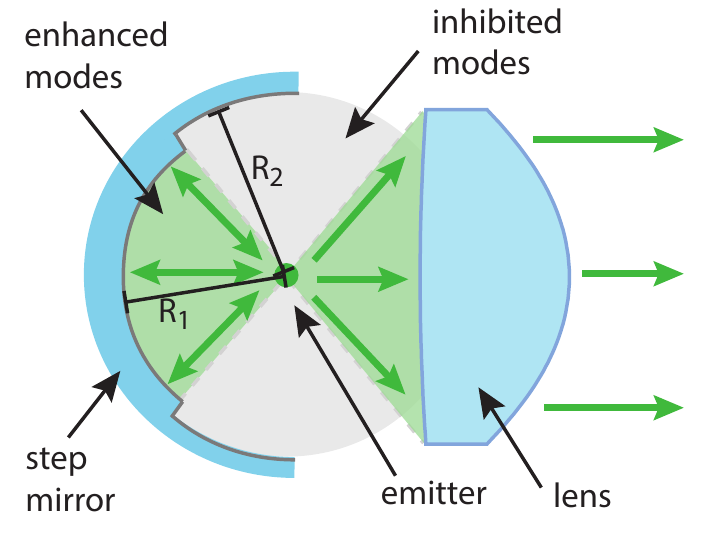}}
\caption[Step mirror concept]{\textbf{Step mirror concept.} A mirror with radius of curvature $R_1=n\lambda/2$ around the center and $R_2=n\lambda/2 +\lambda/4$ in the periphery would enhance the emission of photons in the solid angle collected by the lens while inhibiting the emission of uncollected light.
}
\label{fig:step_mirror}
\end{figure}

The fabrication of macroscopic concave hemispherical mirrors with unprecedented precision will give us access to enhancement and inhibition of the spontaneous emission of a single atom of more than 96\,\% of its free-space value. The fabrication technique could be extended to obtain other concave surfaces, allowing for more exotic quantum electrodynamics situations. One example would be a hemispherical mirror with a $\lambda/4$ radius step as shown in Fig.~\ref{fig:step_mirror} (details can be found in Ref.~\onlinecite{GabrielThesis}). Such a mirror could be used to enhance the spontaneous emission in the modes collected by a lens, while inhibiting the rest, resulting in collection efficiencies close to 100\,\% for a perfectly shaped mirror. For a mirror with a $\lambda/4$ step, fabricated with the surface accuracy demonstrated in section~\ref{sec:mirror}, combined with the $\text{NA} = 0.7$ lens presented in section~\ref{sec:lens}, an unprecedented 68.5\,\% single-mode collection efficiency is expected for a linear dipole oriented perpendicular to the optical axis, in comparison with the 31.7\% expected with a non-stepped mirror~\cite{DanielThesis}.

We have also presented the design of a monolithic high optical access ion trap. Such a trap provides the optical access required for the observation of the quantum electrodynamics effects described above. The trapping parameters are similar to those in state of the art macroscopic traps, such as our ``Innsbruck-style" blade trap~\cite{RotterThesis}. The design of this new trap employs 3D-printing technologies. The basic concept of using trenches to separate electrodes can be extended to more complex trap geometries, including segmented traps or traps designed for trapping of 2D or 3D ion crystals, with high optical access to the trapped particles. The trap fabrication process does not require alignment of different parts nor shadow masking, which makes it reliable and repeatable. The exceptionally high optical access achieved by the trap could also be used in ion setups that integrate an optical cavity~\cite{steiner2013single,brandstatter2013integrated}.

We have also presented the design of the complete setup, which includes a state of the art vacuum vessel and an assembly for fast loading of ions in the trap via laser ablation. With several of these setups one could perform experiments related to remote entanglement distribution and quantum networking without optical cavities~\cite{hucul2015modular,stephenson2019high}.

The physical phenomena that we aim to study with this setup, such as strong inhibition and enhancement of the spontaneous emission, are not restricted to atomic ions, but could be observed from any quantum emitter. The optical setup presented here could be used with other systems with promising prospects in quantum networking and communications, such as quantum dots or diamond spin qubits~\cite{lodahl2017quantum,humphreys2018deterministic}.

\section*{Supplementary Material}
The supplementary material includes a STEP file with the trap design. This file can be opened with any standard CAD software such as SolidWorks, AutoCAD or SALOME. Additionally, we include a 3D PDF with a 3D view of the trap. For a proper view of this file we recommend using Adobe Acrobat Reader . 

\begin{acknowledgments}
We thank Christoph Wegscheider and the rest of our mechanical workshop team for their advice and help with the fabrication of the mechanical components. We gratefully acknowledge the contribution of W. Shihua and colleagues at the National Metrology Centre, A*STAR in Singapore. Without their expertise and the use of their high-NA optical interferometer and contact-probe measurements the first stages of this work would not have been possible. We also thank the FEMTOprint, Elceram and Asphericon teams for their advice and collaboration. We thank Lukas Slodicka, Martin Van Mourik, Pavel Hrmo, Ezra Kassa, Simon Ragg and Chiara Decaroli for fruitful discussions. This work was supported by the European Commission through project PIEDMONS 801285 and by the Institut für Quanteninformation GmbH. This work was supported by the European Commission also through the Marie Skłodowska-Curie Action, Grant Number: No 801110 (Erwin Schrödinger Quantum Fellowship Programme).
\end{acknowledgments}

\section*{Data Availability}
The data that support the findings of this study are available from the corresponding author
upon reasonable request.

%\nocite{*}
%\newpage
\clearpage
%\bibliography{thesis_library.tex}

\end{document}